\documentclass{svproc}
\usepackage{url}
\usepackage{amsmath}

\newcommand{\ud}{\mathrm{d}}
\newcommand{\uvec}[1]{\boldsymbol{#1}}
\newcommand{\LRD}{\overset{\leftrightarrow}{D}\!\!\!\!\!\phantom{D}}

\begin{document}
\mainmatter              
\title{The origin of the nucleon mass}
\titlerunning{Nucleon mass}  
%
\author{C\'edric Lorc\'e\inst{1}}
\authorrunning{C\'edric Lorc\'e} 
%
%
\institute{Centre de Physique Th\'eorique, \'Ecole polytechnique, 
	CNRS, Universit\'e Paris-Saclay, F-91128 Palaiseau, France\\\email{cedric.lorce@polytechnique.edu}}

\maketitle              

\begin{abstract}
It is often claimed that 98\% of the nucleon mass is generated by quantum chromodynamics. The decomposition of the nucleon mass based on the trace of the energy-momentum tensor suggests that gluons play by far a dominant role. About 25 years ago, Ji proposed another decomposition based on the energy component of the energy-momentum tensor, leading to a quite different picture. Recently, we critically revisited these decompositions and argued that both overlooked pressure effects. In particular we showed that Ji's decomposition, although mathematically correct, makes little sense from a physical point of view. We identify the proper mass decomposition along with a balance equation for the pressure forces.
\end{abstract}

\section{Introduction}

While the Brout-Englert-Higgs mechanism generates the current quark masses, it accounts only for about 2\% of the nucleon mass. The remaining 98\% comes from the relativistic kinetic energy and the strong interactions confining quarks and gluons inside hadrons~\cite{Schumacher:2014jga,Gao:2015aax}, described by quantum chromodynamics (QCD).

The question of the origin of the nucleon mass can be addressed based on the QCD energy-momentum tensor (EMT)
\begin{equation}
T^{\mu\nu}=\overline\psi\gamma^\mu \tfrac{i}{2}\LRD^\nu\psi-G^{a\mu\lambda}G^{a\nu}_{\phantom{a\nu}\lambda}+\tfrac{1}{4}\,\eta^{\mu\nu}G^2,
\end{equation}
where $\psi$ is the quark field, $\LRD^\mu=\overset{\rightarrow}{\partial}\!\!\!\!\phantom{\partial}^\mu-\overset{\leftarrow}{\partial}\!\!\!\!\phantom{\partial}^\mu-2igA^{a\mu}t^a$ is the symmetric non-abelian covariant derivative, $G^{a\mu\nu}$ is the gluon field strength, and $\eta_{\mu\nu}$ is the Minkowski metric. The sum over the quark flavors is implicit. Note that in Particle Physics there is no fundamental reason for requiring the EMT to be symmetric~\cite{Leader:2013jra,Lorce:2015lna}. Beside the classical term $\overline\psi m\psi$, the trace of the renormalized EMT
\begin{equation}\label{trace}
T^\mu_{\phantom{\mu}\mu}=\tfrac{\beta(g)}{2g}\,G^2+(1+\gamma_m)\,\overline\psi m\psi,
\end{equation}
includes anomalous quantum contributions involving the $\beta$ function and the anomalous quark mass dimension $\gamma_m$, see e.g.~\cite{Crewther:1972kn,Chanowitz:1972vd,Adler:1976zt,Nielsen:1977sy}.

In the following, we critically review the two standard mass decompositions found in the literature, and propose a new one free of the problems associated with the former~\cite{Lorce:2017xzd,Lorce:2018egm}.

\section{Standard decompositions}\label{sec2}

\subsection{Trace decomposition}

Using the covariant normalization
\begin{equation}\label{statenorm}
\langle P'|P\rangle=2P^0\,(2\pi)^3\,\delta^{(3)}(\uvec P'-\uvec P),
\end{equation}
Poincar\'e invariance imposes that the forward matrix elements of the total EMT in a nucleon state with momentum $P$ take the form~\cite{Jaffe:1989jz}
\begin{equation}\label{param0}
\langle P|T^{\mu\nu}(0)|P\rangle=2P^\mu P^\nu.
\end{equation}
Considering the trace of this expression and using Eq.~\eqref{trace}, one finds that
\begin{equation}\label{tracedec}
2M^2=\langle P|\tfrac{\beta(g)}{2g}\,G^2|P\rangle+\langle P|(1+\gamma_m)\,\overline\psi m\psi|P\rangle.
\end{equation}
Since the second term is known to give a rather small contribution, this picture suggests that most of the nucleon mass comes from gluons~\cite{Shifman:1978zn,Roberts:2016vyn,Krein:2017usp}.

Although manifestly covariant, we find that the physical interpretation of this decomposition is somewhat misleading for the following reasons:
\begin{enumerate}
\item We know from Quantum Mechanics that we are in principle free to choose the normalization of states since physical quantities associated with an operator $O$ are expressed as $\langle\Psi|O|\Psi\rangle/\langle\Psi |\Psi\rangle$. The standard trace decomposition does not involve the normalization factor $1/\langle P|P\rangle$ and appears to be manifestly covariant only because of the particular choice~\eqref{statenorm}.  
\item The relation between the trace of the EMT and the nucleon mass holds only at the level of the matrix elements and for the total EMT. At the operator level, it is knot known how to relate the individual operators $\tfrac{\beta(g)}{2g}\,G^2$ and $(1+\gamma_m)\,\overline\psi m\psi$ to actual gluon and quark contributions to the nucleon mass.
\end{enumerate}
Note that the forward matrix elements of any scalar operator are necessarily proportional to some power of the nucleon mass, since the latter is the only natural scale at our disposal. If we followed the same logic as with the trace operator, many of these scalar operators would lead to quite different ``decompositions'' of the nucleon mass. In summary, although the decomposition~\eqref{tracedec} is mathematically correct, one has to be very careful with the physical interpretation of the individual terms.

\subsection{Ji's decomposition}

A decomposition of the nucleon mass analogous to the virial theorem for a harmonic oscillator and the hydrogen atom has been proposed by Ji~\cite{Ji:1994av,Ji:1995sv}. The idea is to decompose first the renormalized QCD EMT into traceless and trace parts
\begin{equation}\label{barhat}
T^{\mu\nu}=\bar T^{\mu\nu}+\hat T^{\mu\nu}\qquad\textrm{with}\qquad\hat T^{\mu\nu}=\tfrac{1}{4}\,\eta^{\mu\nu}T^\alpha_{\phantom{\alpha}\alpha}.
\end{equation}
The traceless part is then further decomposed into quark and gluon contributions, whereas the trace part is further decomposed into quark mass and trace anomaly contributions
\begin{equation}\label{Tmunudec}
T^{\mu\nu}=\bar T^{\mu\nu}_q+\bar T^{\mu\nu}_g+\hat T^{\mu\nu}_m+\hat T^{\mu\nu}_a.
\end{equation}
The corresponding forward matrix elements can be written as
\begin{align}
\langle P|\bar T^{\mu\nu}_q(0)|P\rangle&=2\,a(\mu^2)\left(P^\mu P^\nu-\tfrac{1}{4}\,\eta^{\mu\nu}M^2\right),\label{Tq}\\
\langle P|\bar T^{\mu\nu}_g(0)|P\rangle&=2\,[1-a(\mu^2)]\left(P^\mu P^\nu-\tfrac{1}{4}\,\eta^{\mu\nu}M^2\right),\\
\langle P|\hat T^{\mu\nu}_m(0)|P\rangle&=\tfrac{1}{2}\,b(\mu^2)\,\eta^{\mu\nu}M^2,\\
\langle P|\hat T^{\mu\nu}_a(0)|P\rangle&=\tfrac{1}{2}\,[1-b(\mu^2)]\,\eta^{\mu\nu}M^2,\label{Ta}
\end{align}
where the coefficients $a(\mu^2)$ and $b(\mu^2)$ depend generally on the renormalization scheme and scale $\mu$. 

According to the Hamiltonian formalism, the mass of a system is identified with the total energy defined in the rest frame. Ji then proposed to decompose the nucleon mass as
\begin{equation}\label{totalmass}
M=M_q+M_g+M_m+M_a,
\end{equation}
where the various contributions on the right-hand side are defined as
\begin{equation}
M_i=\left.\frac{\langle P|H_i|P\rangle}{\langle P|P\rangle}\right|_{\uvec P=\uvec 0}\qquad i=q,g,m,a
\end{equation}
with
\begin{align}
H_q&=\int\ud^3r\,\psi^\dag(i\uvec D\cdot\uvec\alpha)\psi,\label{Hq}\\
H_g&=\int\ud^3r\,\bar T^{00}_g(r),\\
H_m&=\int\ud^3r\left(1+\tfrac{1}{4}\,\gamma_m\right)\overline\psi m\psi,\\
H_a&=\int\ud^3r\,\hat T^{00}_a(r).\label{Ha}
\end{align}
Note that the QCD equations of motion have been used to rearrange quark mass contributions between $\bar T^{00}_q$ and $\hat T^{00}_m$. Using the parametrization in Eqs.~\eqref{Tq}-\eqref{Ta}, one finds that $M_q=\tfrac{3}{4}\left(a-\tfrac{b}{1+\gamma_m}\right)M$, $M_g=\tfrac{3}{4}\left(1-a\right)M$, $M_m=\tfrac{1}{4}\,\tfrac{4+\gamma_m}{1+\gamma_m}\,b\,M$, and $M_a=\tfrac{1}{4}\left(1-b\right)M$.

Sometimes, Ji's decomposition is criticized because it is obtained in the nucleon rest frame~\cite{Roberts:2016vyn}. We do not consider this as an actual problem since most of the physical quantities, like e.g. energy and momentum, are known to be frame-dependent. For massive systems like the nucleon, the rest frame plays a special role and is commonly chosen as the preferred frame for a decomposition. Note also that if one insists on manifest Lorentz invariance, it is always possible to trade a frame-dependent quantity defined in a particular frame for a frame-independent quantity with simple interpretation in that particular frame only~\cite{Hoodbhoy:1998bt,Leader:2013jra}. The archetypical example is the four-momentum squared $p^2=(p^0)^2-\uvec p^2=m^2$, where $m=p^0|_{\uvec p=\uvec 0}$ represents the rest-frame energy. In the case of Ji's decomposition, a covariant form can formally be obtained by trading $\langle T^{00}\rangle|_{\uvec P=\uvec 0}$ for the Lorentz-invariant quantity $\langle T^{\mu\nu}u_\mu u_\nu\rangle$, where $u^\mu=P^\mu/M$ is the nucleon four-velocity.

The actual problem with Ji's decomposition is that although $T^{00}$, $\bar T^{00}$ and $\hat T^{00}$ all have the dimension of energy densities, they correspond to \emph{different} thermodynamic potentials. This is not apparent because the non-covariant treatment, focused on the $\mu=\nu=0$ component in the rest frame, is unable to distinguish pressure-volume work from other forms of energy. In summary, although Ji's decomposition~\eqref{totalmass} is also mathematically correct, it amounts to adding apples and oranges since all four individual contributions correspond to four different combinations of internal energy and pressure-volume work.

\section{New decomposition}\label{sec3}

Lorentz covariance implies that the mass decomposition follows directly from a decomposition of the EMT. Since the various components of the EMT correspond to different mechanical properties, one should not consider a decomposition based on the tensor structure like in Eq.~\eqref{barhat}, but rather a decomposition based on the sole nature of the constituents. The QCD EMT can naturally be decomposed into quark ($i=q$) and gluon ($i=g$) contributions $T^{\mu\nu}=\sum_iT^{\mu\nu}_i$. The corresponding forward matrix elements in a nucleon state read~\cite{Ji:1996ek,Leader:2013jra}
\begin{equation}\label{param}
\langle P| T^{\mu\nu}_i(0)|P\rangle=2P^\mu P^\nu A_i(0)+2M^2\eta^{\mu\nu}\bar C_i(0),
\end{equation}
where $A_i(0)$ and $\bar C_i(0)$ are two gravitational form factors (GFFs) evaluated at zero momentum transfer. Poincar\'e invariance~\eqref{param0} implies that $\sum_iA_i(0)=1$ and $\sum_i\bar C_i(0)=0$. Defining the quark and gluon contributions as 
\begin{equation}\label{qg}
T^{\mu\nu}_q\equiv\bar T^{\mu\nu}_q+\hat T^{\mu\nu}_m\qquad\text{and}\qquad T^{\mu\nu}_g\equiv\bar T^{\mu\nu}_g+\hat T^{\mu\nu}_a,
\end{equation}
these GFFs can easily be related to Ji's coefficients as follows $a_i=A_i(0)$, $b_i=A_i(0)+4\,\bar C_i(0)$.

If we average the expectation value of $T^{\mu\nu}_i$ over the nucleon volume $\mathcal V= V\,M/P^0$
\begin{equation}
\langle\langle T^{\mu\nu}_i\rangle\rangle\equiv\frac{1}{\mathcal V}\,\frac{\langle P|\int\ud^3r\,T^{\mu\nu}_i(r)|P\rangle}{\langle P|P\rangle}=\frac{\langle P|T^{\mu\nu}_i(0)|P\rangle}{2M^2}\,\frac{M}{V},
\end{equation}
we find using $u^\mu=P^\mu/M$~\cite{Lorce:2017xzd,Lorce:2018egm}
\begin{equation}\label{expT}
\langle\langle T^{\mu\nu}_i\rangle\rangle=(\varepsilon_i+\bar p_i)\,u^\mu u^\nu- p_i\,\eta^{\mu\nu}
\end{equation}
with
\begin{equation}\label{ep}
\varepsilon_i=[A_i(0)+\bar C_i(0)]\,\tfrac{M}{V}\qquad\text{and}\qquad p_i=-\bar C_i(0)\,\tfrac{M}{V}.
\end{equation}
The structure in Eq.~\eqref{expT} is similar to that of an element of perfect fluid in relativistic hydrodynamics and allows us to interpret $\varepsilon_i$ and $p_i$ as partial proper internal energy density and isotropic pressure averaged over the nucleon, respectively. Both are expressed in Eq.~\eqref{ep} in units of the average density $M/V$. 

Multiplying $\varepsilon_i$ and $p_i$ by the nucleon proper volume $V$, we obtain the partial internal energy and pressure-volume work
\begin{equation}
U_i=\varepsilon_iV=[A_i(0)+\bar C_i(0)]\,M,\qquad W_i=p_iV=-\bar C_i(0)\,M
\end{equation}
which satisfy the sum rules
\begin{equation}\label{sumrule}
M=\sum_i U_i,\qquad 0=\sum_i W_i
\end{equation}
derived from Poincar\'e invariance~\eqref{param0}. The first sum rule is nothing but the mass decomposition we were looking for. The second sum rule expresses the stability of the nucleon by imposing that the total pressure forces must balance between the various parts of the system.

\section{Discussion}

Now that internal energy and pressure-volume work contributions are well identified, we can unravel the meaning of the old decompositions. Starting with the trace decomposition~\eqref{tracedec} divided by $2M$, we see that it does not correspond to a decomposition of the total energy of the system, but rather to a decomposition of the interaction measure $I=\sum_i I_i$ with $I_i=U_i-3W_i$. Since the total pressure-volume work vanishes, the total interaction measure coincides with the nucleon mass $I=M$. The fact that the gluon contribution dominates $I_g\gg I_q$ does not mean that it is responsible for most of the nucleon mass as claimed e.g. in~\cite{Shifman:1978zn,Roberts:2016vyn,Krein:2017usp}. It turns out in fact that the nucleon mass is more or less equally shared between quarks and gluons $U_q\approx U_g$~\cite{Lorce:2017xzd,Lorce:2018egm}. What the dominance of the gluon contribution to the interaction measure indicates is that the gluon average pressure $p_g=-p_q$ is large and negative. In average, gluons are therefore responsible for the net attractive forces inside the nucleon, balanced by the net repulsive forces associated with quarks.

Turning now to Ji's decomposition, we find that using the definition of quark and gluon EMT given in Eq.~\eqref{qg}, the four terms in Eq.~\eqref{totalmass} correspond to the following combinations of average internal energy and pressure-volume work
\begin{align}
M_q&=\tfrac{3}{4}\,\tfrac{1}{1+\gamma_m}\left[\gamma_m U_q+(4+\gamma_m)W_q\right],\\
M_m&=\tfrac{1}{4}\,\tfrac{4+\gamma_m}{1+\gamma_m}\left(U_q-3W_q\right),\\
M_g&=\tfrac{3}{4}\left(U_g+W_g\right),\\
M_a&=\tfrac{1}{4}\left(U_g-3W_g\right).
\end{align}
Clearly, each term $M_i$ represents a different physical quantity. One is adding apples and oranges, which is not something we would like for a genuine mass decomposition. In order for Ji's decomposition to make sense, one has to give up something. If one sacrifices covariance, one can impose from the onset the further decompositions $U_q=M_q+M_m$ and $U_g=M_g+M_a$, keeping the pressure-volume works unchanged. If one wants to preserve covariance, one then has to treat $\bar T^{\mu\nu}_i$ and $\hat T^{\mu\nu}_i$ as actual separate EMTs. In that case, one is fixing arbitrarily the equation of state of the individual contributions because of the restriction on the Lorentz structure. For example, the gluon contribution in Ji's decomposition is divided into kinetic+potential energy treated as a pure radiation $\bar p_g=\tfrac{1}{3}\,\bar\varepsilon_g$, and trace anomaly treated as a cosmological constant $\hat p_a=-\hat\varepsilon_a$. In other words, a covariant decomposition into four terms can only be achieved by combining in an arbitrary way the two a priori independent sum rules~\eqref{sumrule} into a single one.

\section*{Acknowledgement}

This work is a result of discussions held at the workshop ``The Proton Mass: At the Heart of Most Visible Matter'' at the ECT* Trento, on 3-7 April 2017. It has been supported by the Agence Nationale de la Recherche under the project ANR-16-CE31-0019.

%
%

\end{document}